\documentclass[12pt,preprint]{aastex}

\slugcomment{Not to appear in Nonlearned J., 45.}

\shorttitle{SN with ``Non-Uniform'' Magnetic Field}
\shortauthors{Sawai, Kotake, \& Yamada}

\begin{document}

\title{The Core-Collapse Supernova \\with ``Non-Uniform'' Magnetic Fields}

\author{Hidetomo Sawai\altaffilmark{1}, Kei Kotake\altaffilmark{1} and
Shoichi Yamada\altaffilmark{1,2}}
\altaffiltext{1}{\textit{Science \& Engineering, Waseda University,
3-4-1 Okubo,
Shinjuku, Tokyo 169-8555, Japan}}
\email{hsawai@heap.phys.waseda.ac.jp}

\altaffiltext{2}{\textit{Advanced Research Institute for Science \&
Engineering, Waseda University, 3-4-1 Okubo,
Shinjuku, Tokyo 169-8555, Japan}}

\begin{abstract}
 We perform two-dimensional numerical simulations on the
 core-collapse of a
 massive star with strong magnetic fields and differential
 rotations using a numerical code ZEUS-2D. Changing field configurations
 and laws of differential 
 rotation parametrically, we compute 14 models and investigate
 effects
 of these parameters on the dynamics. In our models, we do not solve the
 neutrino transport 
 and instead employ a phenomenological parametric EOS that takes into
 account the neutrino emissions. As a result of the
 calculations, we find that the field configuration plays a significant
 role in the dynamics of the core if the initial magnetic field is large
 enough. Models with initially concentrated fields produce more energetic
 explosions
 and more prolate shock waves than the uniform
 field. Quadrapole-like fields produce remarkably collimated and fast
 jet, which might be important for gamma-ray bursts(GRB). The Lorentz
 forces exerted in the region where 
 the plasma-beta is less than unity are responsible for these dynamics.
 The pure toroidal field, on the other hand, does not lead to any explosion or
 matter ejection. This suggests the presupernova models of
 \citet{heg03}, in which toroidal fields are predominant, is
 disadvantageous for the
 magnetorotation-induced supernova considered here. Models with
 initially weak magnetic fields do not lead to
 explosion or matter ejection, either. In these models magnetic
 fields play no role as they do not grow on the timescale considered in
 this paper so that the magnetic
 pressure could be comparable to the matter pressure. This is because the
 exponential field growth as expected in MRI is not seen in our
 models. The magnetic
 field is amplified mainly by field-compression and field-wrapping in our
 simulations. 
\end{abstract}

\keywords{supernovae: general --- stars: magnetic fields --- pulsars:
general --- magnetars: general --- MHD --- methods: numerical  }

\section{Introduction}
The study of magnetorotational core-collapse supernovae is currently attracting
great attention. This is mainly because 
anomalous X-ray pulsars (AXP) \citep{tho96} and soft gamma-ray
repeaters (SGR) \citep{dun92} have been
discovered and thought to be candidates of magnetar. The magnetar is a
sub-class of pulsar which has
an extraordinarily large magnetic field, $\sim$ $10^{14}$ - $10^{15}$
gauss. This value is two to three orders of magnitude greater than that
of the ordinal
pulsar. So far, only about ten of them have been observed and little is
known of them. The formation mechanism, in particular, is
veiled in mystery.   

Ordinary pulsars are thought to be
formed as a result of core-collapse supernovae, and so are magnetars. Then, it
is necessary to study magnetorotational core-collapse supernovae. Since
the number of magnetars is smaller than
that of ordinary pulsars by a factor of $\sim 100$, they form a
particular group and there may be a special
condition for a progenitor, such as large magnetic field or rapid
rotation, for their formation. The main purpose in this paper is to
systematically study the dynamics
of core-collapse with very strong magnetic fields and rapid rotations.
It may be possible that the study
of this extraordinary supernovae have some implication also for the
ordinary supernovae mechanism which produces normal pulsars. 

The mechanism of ordinary supernovae is still
unknown. The recent spherically symmetric simulations which employ a
realistic equation of 
state (EOS) and sophisticated microphysics 
such as the neutrino transport and/or the electron
capture \citep[][]{ram00, lie01, tho03, bur03, lie04, sum04} have not
found successful explosions. On the
other hand, core-collapse supernovae might be generically
asymmetric \citep{wan96, leo00}. SN1987A is a clear example,
which is indicated by the HST image of the asymmetrically expanding
envelope \citep[][and references therein]{whe00}. Rotation of
progenitors is a natural choice for the cause of asymmetry although the
instability of the standing accretion shock may be another candidate
\citep{fog00, blo03}. Some authors claimed that rotation and asymmetric
neutrino radiation induced thereby may be crucially important for the
explosion \citep[][but see also \citet{wal05} and \citet{jan05} for
critieisms]{shi01, kot03, yam05}. On the other hand magnetic fields in
massive stars might yet another candidate for the cause of not only
asymmetry but also explosion itself. In fact, some researchers are
attempting to explain the
mechanism for all supernovae with magnetic field \citep{whe02}. 

The first numerical simulation of supernovae with magnetic field
and rotation was done about thirty years ago by
\citet{leb70}. Some more numerical studies followed them after
\citep[e.g.][]{ bis75, mul79, ohn83, sym84}. Although these studies, especially
\citet{sym84}, are important, they had not attracted much 
attention mainly because there were no observational support that 
magnetic fields may play an important role in supernovae one way or
another. In fact, the field strength inferred from the ordinary pulsar is
negligible for the dynamics of core-collapse. 
The situation, however, may have 
changed with the discovery of magnetars. The progress of our
understanding of the mechanism for field amplification \citep{bal91} is
yet another boost. In the last couple of years, we have seen the study
of magnetized supernovae have gained momentum
again \citep[e.g.][]{ard00, whe02, aki03, yam04, kot04, tak04, ard04}.

The number of numerical models, however, is still not very large. Even
the systematics of dynamics for magneto-rotational core-collapse has not
been investigated throughly. We
studied effects of strong uniform, poloidal magnetic field with rapid
rotation systematically, varying the initial field strengths as
parameters \citep{yam04}. It was found that
the jet-like explosion is produced by the combination of initial large
magnetic field and rapid rotation and that the
driving force is the amplified magnetic fields in the region between
the shock wave and the inner core. In this paper, only the uniform field
was considered as an initial field configuration. In fact, the effect of
the initial field configuration has not been studied systematically so
far. It is true that the initially uniform field does not have a
firm basis. As a matter of fact, recent studies of stellar evolution
\citep{heg03} suggest that the toroidal fields are dominant prior to the
collapse. This is, however, still highly uncertain. Hence, in this paper
we investigate how dynamics and field amplifications depend on the
initial field configurations, assuming them rather arbitrarily. As
mentioned later again in \S \ref{bandr}, we
mainly explore a very strong field regime, in which we assume
$B\sim10^{12} - 10^{13}$ G initially.
However, we also study weak field models with $B\sim10^{8} - 10^{9}$ G
for comparison. Our standing point is that we are concerned with the
strongly magnetized progenitors, which will produce magnetars and we do
not address the origin of such strong magnetic fields for the
moment. Since the main purpose is to study the systematics of dynamics,
we simplify microphysics and study the phenomena occurring only on the 
prompt-explosion timescale as in \citet{yam04}. The
present paper is a sequel of \citet{yam04}.

The paper is organized as follows. We introduce our methods of
calculation and models in the next 
section. The results are presented in \S 3. In the last section
we discuss our results and conclude this paper.

\section{Numerical Methods and Models}
  
\subsection{Numerical Code}
We have carried out two-dimensional axisymmetric magnetohydrodynamic
(MHD) simulations with
the numerical code ZEUS-2D developed by \citet{sto92}. We describe some
properties of the code briefly. There are two main
difficulties in solving the MHD equations compared to the hydrodynamic
(HD) equations. The
first one is to deal with the constraint of the magnetic field, and the second
one is concerned with the accurate treatment of Alfv\'en waves. For the first
difficulty, ZEUS-2D employs the constrained transport (CT) method instead
of solving the vector potential, which would produce false
accelerations and heating near shocks or contact surfaces. As for the
second problem, the method of characteristics (MOC) is 
employed in order to avoid incorrect Alfv\'en modes. In solving the Poisson
equation for the gravitational potential, this code utilizes the
Incomplete Cholesky decomposition Conjugate Gradient (ICCG) method. Readers are
referred to their original paper \citep{sto92} for more detail.

\subsection{Basic Equations}
We solve the ideal MHD equations,
\begin{eqnarray}
  \frac{D\rho}{Dt}+\rho\nabla\cdot\mbox{\boldmath $v$}=0\label{mac},\\
  \nonumber\\
  \rho\frac{D\mbox{\boldmath $v$}}{Dt}=-\nabla p-\rho\nabla\Phi+\frac{1}{4\pi}
   (\nabla\times\mbox{\boldmath $B$})\times\mbox{\boldmath $B$}\label{moc},\\
  \nonumber\\
  \rho\frac{D}{Dt}(\frac{e}{\rho})=-p\nabla\cdot
   \mbox{\boldmath $v$}\label{enc},\\
  \nonumber\\ 
  \frac{\partial\mbox{\boldmath $B$}}{\partial t}=\nabla\times
   (\mbox{\boldmath $v$}\times\mbox{\boldmath $B$}),\label{far}
\end{eqnarray}
where $\rho$, \mbox{\boldmath $v$}, $p$, $e$, $\Phi$, \mbox{\boldmath
$B$} are the density,
velocity, internal energy density, gravitational potential,
and magnetic
flux density\footnote{Hereafter, it is called magnetic field for the
convenience.}, respectively.
The Lagrangian derivative is denoted as \mbox{$\frac{D}{Dt}$}.

Assuming the equatorial symmetry, we use the spherical
coordinates and solve only the quarter of the meridional plane. Until the
central density reaches $10^{12}$ g/cm$^3$, we use 200 ($r$)$\times$ 30
($\theta$) grid points, extending to 2000 km in the
radial direction. Thereafter, the number of grid points and the radius
of outer boundary are set to be 300 ($r$)$\times$ 30 ($\theta$) and 1500
km, respectively. In the radial direction, the mesh is non-uniform with
finer grids toward the center while the angular grid points are uniform.

\subsection{Equation of State}\label{eossec}   
As in the previous paper of \citet{yam04}, we adopt a parametric EOS
which was first introduced by \citet{tak84}. Since our purpose is to
investigate
effects of the magnetic field configuration on the prompt-explosion
timescale and we are mainly concerned with the systematic change of
dynamics, we drastically simplify complicated microphysics such as the
neutrino transport.

The parametric EOS we employed in this paper is described as follows;
\begin{eqnarray}         
 p_{tot}=p_c(\rho)+p_t(\rho,e_t),\label{eosm}\\
 p_c(\rho)=K\rho^\Gamma,\label{eosc}\\ 
 p_t(\rho,e_t)=(\gamma_t-1)\rho\epsilon_t\label{eost}.
\end{eqnarray}
The pressure consists of two parts, the cold part ($p_c$) and the thermal
part ($p_t$). The thermal part is a function of the density and the
specific thermal energy, $\epsilon_t$, in which $\gamma_t$ is the
parameter called
the thermal stiffness. The thermal energy generated by shock dissipation
loses its considerable part due to the photodisintegration of neuclei as
well as to
thermal neutrino emissions. This effect is mimicked in
the thermal part of EOS with an appropriate value of the thermal
stiffness. On the other hand, the cold part is a function
of the density alone where the constants $K$ and $\Gamma$ reflect the
effect of the degeneracy of
leptons and the nuclear force. The values
of $\Gamma$ are given as follows;
\begin{eqnarray}         
 \Gamma_1=\frac{4}{3} \hspace{1cm}\textrm{dencity regime \hspace{0.3cm} I}\\
 \Gamma_2=\frac{\log p_2-\log p_1}{\log\rho_2-\log\rho_1}=\frac{4}{3}+\frac{\log d}{\log(\rho_2/\rho_1)}\hspace{3.45cm}\textrm{II}\\ 
 \Gamma_3=\frac{4}{3}\hspace{3.45cm}\textrm{III}\\
 \Gamma_4=2.5\hspace{3.45cm}\textrm{IV}\\
 d\equiv \frac{p_2}{p_1}\sim\Biggl[\frac{Y_l(\rho_2)}{Y_l(\rho_1)}\Biggr]^{4/3}\hspace{3.5cm}
\end{eqnarray}
where $Y_l$ is the number of leptons per baryon. The boundary between
the regimes 
I and II corresponds to the onset of the electron capture at density of
$\rho_1 = 4.0\times10^9$cm/g$^3$, from which point the adiabatic index
decreases. The density $\rho_2 = 1.0\times10^{12}$ g/cm$^3$, the boundary
of the regimes II and III, is the point at which the neutrino
trapping is commenced. Then the
electron capture ceases and the adiabatic index increases again. After
the density reaches $\rho_3 = 2.8\times10^{14}$ g/cm$^3$, the pressure becomes
nuclear-force-dominant and adiabatic index rises considerably. In this EOS,
we can specify two parameters, the thermal stiffness $\gamma_t$ and the
lepton fraction $d$. According to the papers by \citet{tak84} and \citet{yam94}
which also used this EOS, larger $\gamma_t$ and $d$ are
favorable for successful prompt explosions. Here we adopt 1.3 for thermal
stiffness and 0.78 for lepton fraction by setting $Y_l(\rho_2)=0.35$
which corresponds to 1.29 for the adiabatic index in the density regime
II. Note that recent sophisticated spherically symmetric simulations
suggest that the
lepton fraction at the onset of neutrino trapping is $\sim 0.35$
\citep[e.g.][]{lie04}. With these parameters, our spherically
symmetric model does not lead to a successful explosion as in recent realistic
simulations.
 
\subsection{Progenitor}
We use the central $1.4M_\odot$ core of
the $15M_\odot$ presupernova stellar model by \citet{woo95}, which
provides us with the spherically symmetric
profile of the density and the specific internal energy. We add by hand
magnetic field
and rotation to the original model, which will be explained in detail in
\S \ref{bandr}. 

\subsection{Magnetic field and Rotation}\label{bandr}
The most important ingredients in this study are magnetic field and
rotation. Changing these parameters, we have computed 14
models. We adopt four different types of field
configurations as follows. The first one, which
is the simplest and has been employed in most of the past simulations, is
uniform field parallel to the rotation axis. The second
configuration is the one which is parallel to the rotation axis but
axially concentrated, and is described as
\begin{equation}
B_z=B_{z0}\frac{\varpi_0^2}{\varpi_0^2+\varpi^2}
\label{con}
\end{equation}
where z and $\varpi$ are the cylindrical coordinates, and
$B_{z0}$ and $\varpi_0$ are constants. One
can obtain strong concentration of the field near the axis with small
$\varpi_0$. The third is
quadrapole-like configuration, which was introduced by \citet{ard98}: 
\begin{eqnarray}
B_{\varpi 0}=F_\varpi(0.5\varpi,0.5z-2.5)-F_\varpi(0.5\varpi,0.5z+2.5), B_{\phi 0}=0,\nonumber\\
B_{z0}=F_z(0.5\varpi,0.5z-2.5)-F_z(0.5\varpi,0.5z+2.5),\\
F_\varpi(\varpi,z)=k\Biggl(\frac{2\varpi z}{(z^2+1)^3}-\frac{2\varpi^3z}{(z^2+1)^5}\Biggr),    
F_z(\varpi,z)=k\Biggl(\frac{1}{(z^2+1)^2}-\frac{\varpi^2}{(z^2+1)^4}\Biggr),\nonumber
\end{eqnarray}
where $k$ is the parameter specifying the strength of magnetic field,
and $z$ and $\varpi$ are normalized by $1.6\times10^8$cm. The last type
is a pure toroidal configuration which is concentrated with the same
law as differential rotation (see below in the text), that is, 
\begin{equation}
B_\phi=B_{\phi 0}\frac{r_0^2}{r_0^2+r^2},
\label{tro}
\end{equation}
where $r$ is the distance from the center, and $B_{\phi 0}$ and $r_0$ are
constants.

We adopt shell-type differential rotation with the
angular velocity distribution,
\begin{equation}
\Omega(r)=\Omega_0\frac{R_0^2}{R_0^2+R^2},
\label{dif}
\end{equation} 
where $\Omega_0$ and $R_0$ are constants. With small $R_0$ strong
differential rotation is obtained.

In Table \ref{models}, we present 14 models. The name of each
model consists of two parts, alphabets and a number. The capital
alphabet denote the
magnetic field configuration with H, C, Q, T representing the
homogeneous, concentrated, quadrapole-like, and toroidal fields,
respectively. And the attached
number stands for a strength of the field concentration and/or
differential rotation. The last three models have a small alphabet (w)
implying that its initial magnetic field is 
very weak. The energy of magnetic field and rotation is set to be
0.5\% of the gravitational energy for all models
except for the last three ones with w in the name. 

According to the recent study on the stellar evolution, a
presupernova
core may be toroidal-field-dominant and may have a slow rotation velocity
\citep{heg03}, but
there still exist some uncertainty in their models, and the distributions of
magnetic field and angular velocity are not well established
yet. Observations are not helpful either. Hence our stand point is that
we regard these distributions as unknown factors and vary them
arbitrarily. If anything, however, model T5w is rather close to that of
\citet{heg03} at least in
the field configuration and strength though the angular velocity
is larger.

\section{Results}\label{result}
In this section we describe the numerical results of the computation. We
mainly 
focus on the differences in dynamics and field-amplification among
the models. In table \ref{parameter}, we show important parameters for
all models.

\subsection{Dynamics}\label{dyn}
\begin{figure}
\plotone{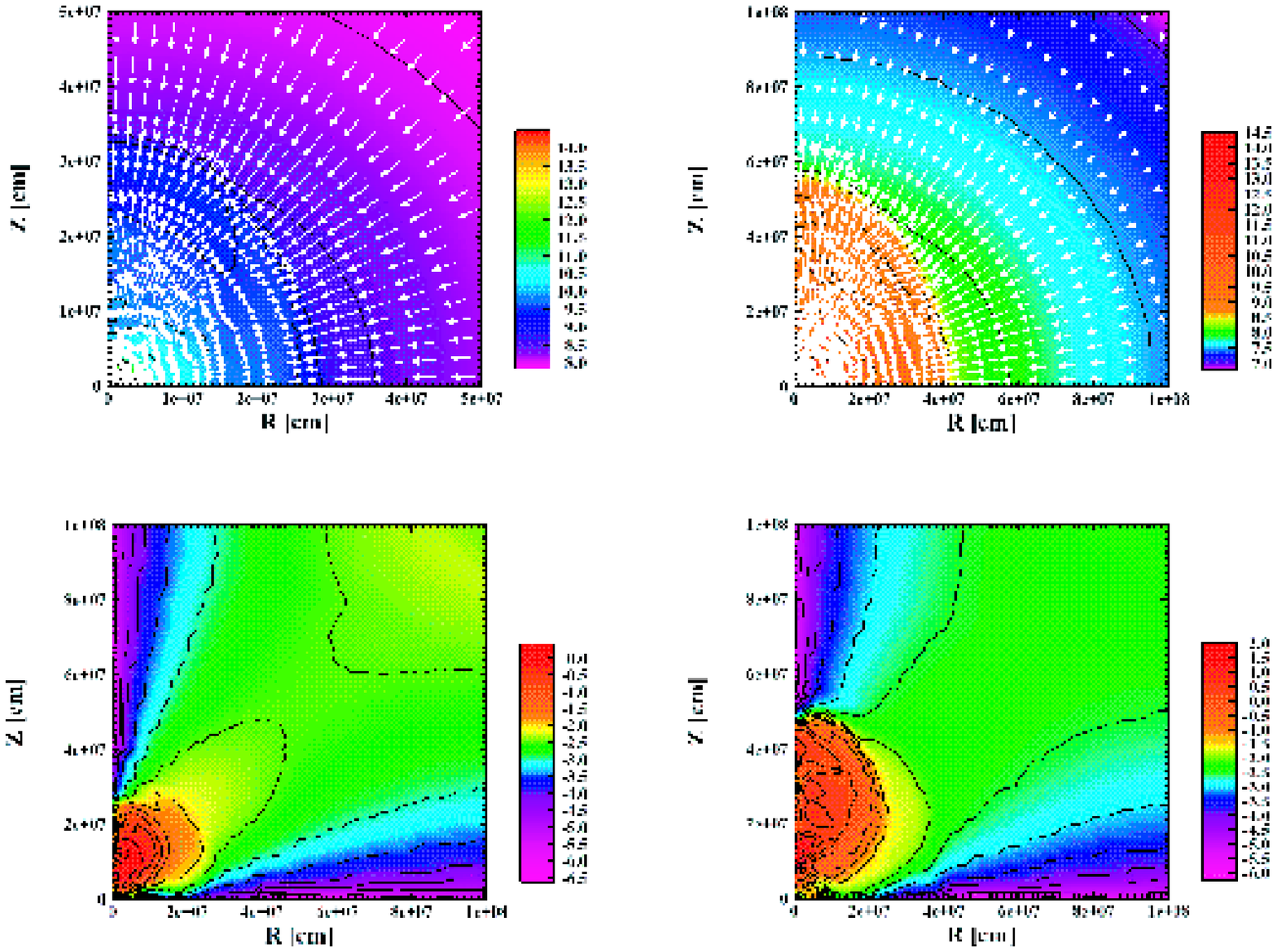}
  \caption{The velocity fields on top of the density color contours for model
 H10 (top panels), and the contours of the ratio of
 $p_{mag}/p_{matter}$, where $p_{mag}$ and $p_{matter}$ are magnetic and
 matter pressures, respectively (bottom two panels). The left figures are
 drawn for at 14 ms after bounce (152 ms from the beggining) and the right
 figures for 24 ms after bounce (162 ms from the begining).}  
  \label{h10}
\end{figure}

\begin{figure}
\begin{center}
\resizebox{10cm}{!}{\includegraphics{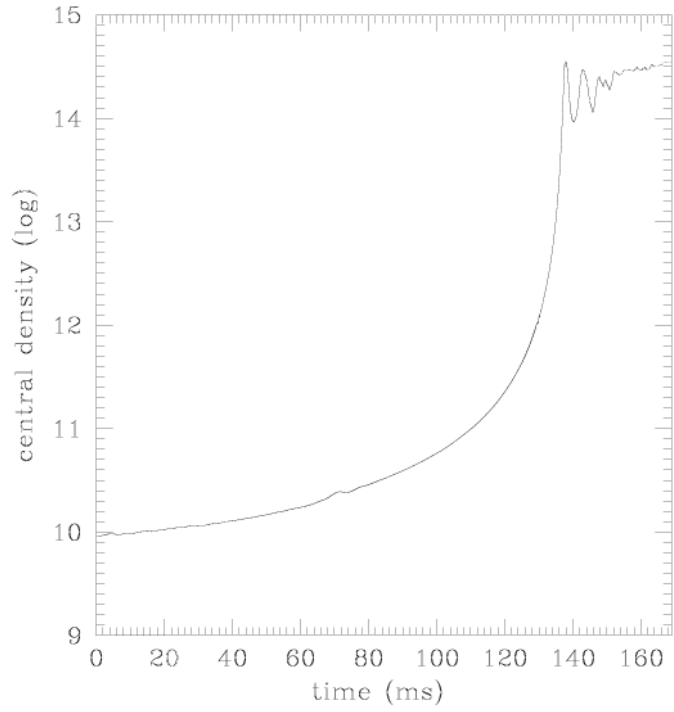}}
\end{center}
  \caption{The time evolution of the central density for model H10.}
  \label{dch10}
\end{figure}

\begin{figure}
\plotone{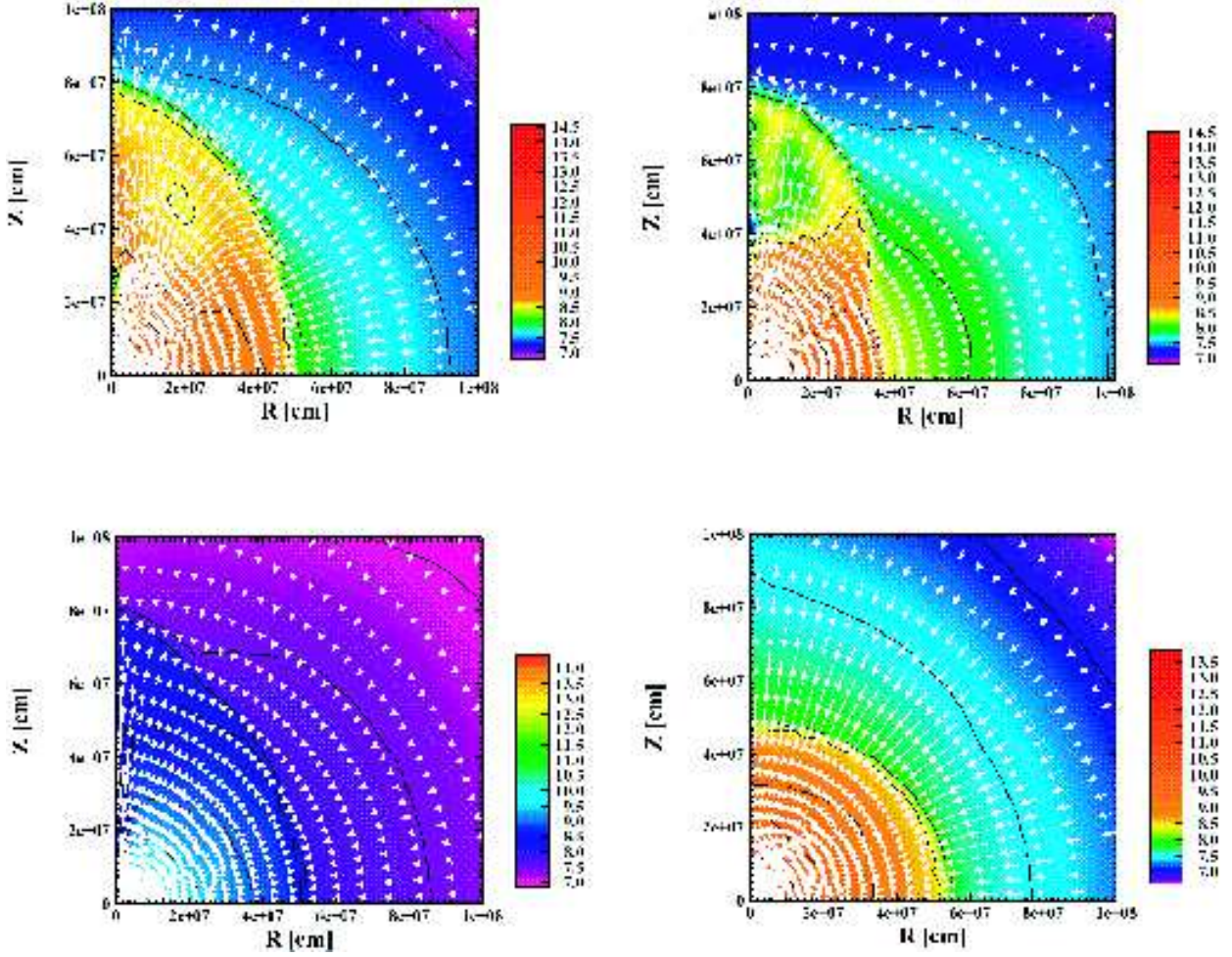}
  \caption{The velocity fields on top of the density contours for model
 H10 (top left panel), C5 (top right panel), Q10 (bottom left panel) and
 T5 (bottom right panel). Each panel is depicted for 31 ms, 27 ms, 42 ms
 and 49 ms after bounce (169 ms, 171 ms, 180 ms and 191 ms from the
 beggining), respectively.}
  \label{veloc}
\end{figure}

We chose model H10 as the reference cases and first describe its
dynamical evolution for the comparison with
other models. The collapsing matter is halted and bounces at 137 ms
after the beginning of simulation,
when the central density reaches its maximum value,
$3.5\times10^{14}$ g/cm$^3$, and a shock wave is produced. During this
collapsing phase, the compression and wrapping of frozen-in magnetic
fields generates a strong toroidal field,
$\sim 10^{16}$ G. A region where the toroidal magnetic pressure
dominates over
the matter pressure begins to be formed behind the shock wave a few
milliseconds after bounce and prevails along the rotation axis as the shock
propagates in a prolate fashion (see Fig. \ref{h10}). As seen in
Fig. \ref{dch10}, the core is 
oscillating for sometime after bounce, and what we call ``small bounces''
produce some more
shock waves. Since these trains of newborn shocks are further powered by
the dominant toroidal magnetic pressure, they gain larger amplitude
in the direction of the rotation axis than the first shock wave which is
not affected by magnetic pressure strongly and does not have enough
energy to penetrate through the core. The first shock is soon catched up
with by these large-amplitude waves and acquires sufficient energy to
break through the core. Consequently, 
a strong magnetocentrifugal jet is induced along the rotation axis (see
Fig. \ref{veloc}), and it will eventually lead to an axisymmetric
bipolar supernova explosion.

On the contrary, no explosion occurs in the pure rotation case (model
R10). The bounce occurs at 142 ms with the
central density, $2.0\times10^{14}$ g/cm$^3$. Small bounces
occurs also in this model. Lacking in the support by magnetic fields,
however, the shocks launched by the small
bounces have too small energy to supply energy to the first
shock when they catch up with it. Note that the synergically symmetric
model does not give an explosion with the current set of the parameters of
our EOS.     

For the models with initially axially-concentrated magnetic fields, the
dynamics is a
little different from that of reference model with the uniform field.
For model C5, in particular, 
the first shock is powered not by the nuclear force but by the magnetic
force. Slightly prior to the nuclear-force-induced bounce, the
magnetic-force-dominant region
has already been formed at the boundary of the inner and outer
cores. Then the first shock wave is generated and starts to
propagate outward. However, it slows down as it goes out of
the region where the magnetic pressure is dominant. The
second shock generated by nuclear force runs after the first shock
and is powered in the magnetic-pressure-dominant region. Subsequent
shock waves generated by ``small bounce'' also gain
large energy in the same way. The first shock soon collects energy
from these
shock waves as in the reference model.  At the end of the
simulation, a strong axial jet is formed with velocities higher than in
model H10. While the jet-collimation is not very different from that of model
H10, the shape of shock surface is more prolate with an aspect ratio of
2.0 (see Fig. \ref{veloc}). 

For model C10, no shock wave is generated prior to the
nuclear-force-induced bounce as in
model C5. The shock wave generated by the bounce, however, acquires large
energy from subsequent shock waves as in the case of uniform field. The
resulting dynamical feature is 
almost the same as in model C10 though the jet collimation is slightly
weaker than
that of model H10 or C5 (see Fig. \ref{veloc}).

As shown in
Table \ref{parameter}, the stronger the field concentration and
differential rotation become, the more prolate the shock is generated
and the faster the jet becomes. According to
\citet{yam94}, strong differential rotation tends to enhance asymmetry
of shock. The comparison of models H10 and C10, which have different
concentration of magnetic fields, suggest that the field
concentration also tends to make a shock more prolate. The reason is
that strong poloidal fields prevent matter from traveling in the transverse
direction. 

The models with the quadrapole-like configuration have distinct
feature, that is, a fast jet and its remarkable collimation with a very small
opening angle\footnote{We measure the collimation of the jet by the
angle $\Delta \theta$ between rotation axis and the point on the shock
surface whose expansion velocity is half the maximum value on the
shock. The angular resolution put the minimum opening angle to be
$6^\circ$ in our simulations.} (see Table \ref{parameter} and
Fig. \ref{veloc}), while the
shock surfaces are less prolate than in the uniform or axially-concentrated
field cases. In these models both the first shock and subsequent shocks
are accelerated toward the rotation axis by the dominant toroidal magnetic
pressure. In model Q10, the 
fastest jet among all models is produced\footnote{The opening
angle for this model is $6^\circ$, which is equal to the angular
resolution. In order to verify that the narrow jet is not a numerical
artifact, we calculate the same model 
with a doubled angular resolution, that is, 60 mesh points in the
$\theta$ direction, and find the same opening angle.}. In this model,
the acceleration
is very strong especially for a subsequent shock, which
is born at 100 km from the center a few ten milliseconds after
bounce when the first shock reaches around 500 km. Then this shock soon gains
large energy from magnetic pressure, and the matter velocity becomes
considerably high, $c/3 - c/2$, at the shock front, where $c$ is the light
velocity. This large amplitude
shock overtakes the first shock and causes the remarkable
collimation. We
evaluate the parameters in Table \ref{parameter} when the shock reaches
800 km from the center on the rotation axis. The large-amplitude
subsequent shock has not caught up with
the first shock yet in model Q10. Although the collimation parameter
for Q10 is not small compared with
models Q3 or Q5, it will become smaller and comparable to
model Q3 later.

In the pure toroidal field cases, we cannot find any substantial deviation
from the rotation-only case though they have large magnetic energy.
Fig. \ref{veloc} shows for model T5 that the
shock wave stalls around 400 km, and neither explosion nor matter
ejection occurs. Moreover, no region appears where the magnetic
pressure is dominant through the simulation. This
is because the field-wrapping by differential rotation does not
occur. It is true that the compression of frozen-in fields during the
collapse amplifies the toroidal fields significantly 
but the field grows as $B \propto \rho^{3/4}$ which
is the same as the increase of matter pressure. Hence the ratio of
magnetic pressure 
to matter pressure is unchanged. Note that \citet{kot04} found
magnetic-field-dominant regions are formed in their pure toroidal
models, which may be ascribed the difference in employed EOS and
including microphysics.

\begin{figure}
\plotone{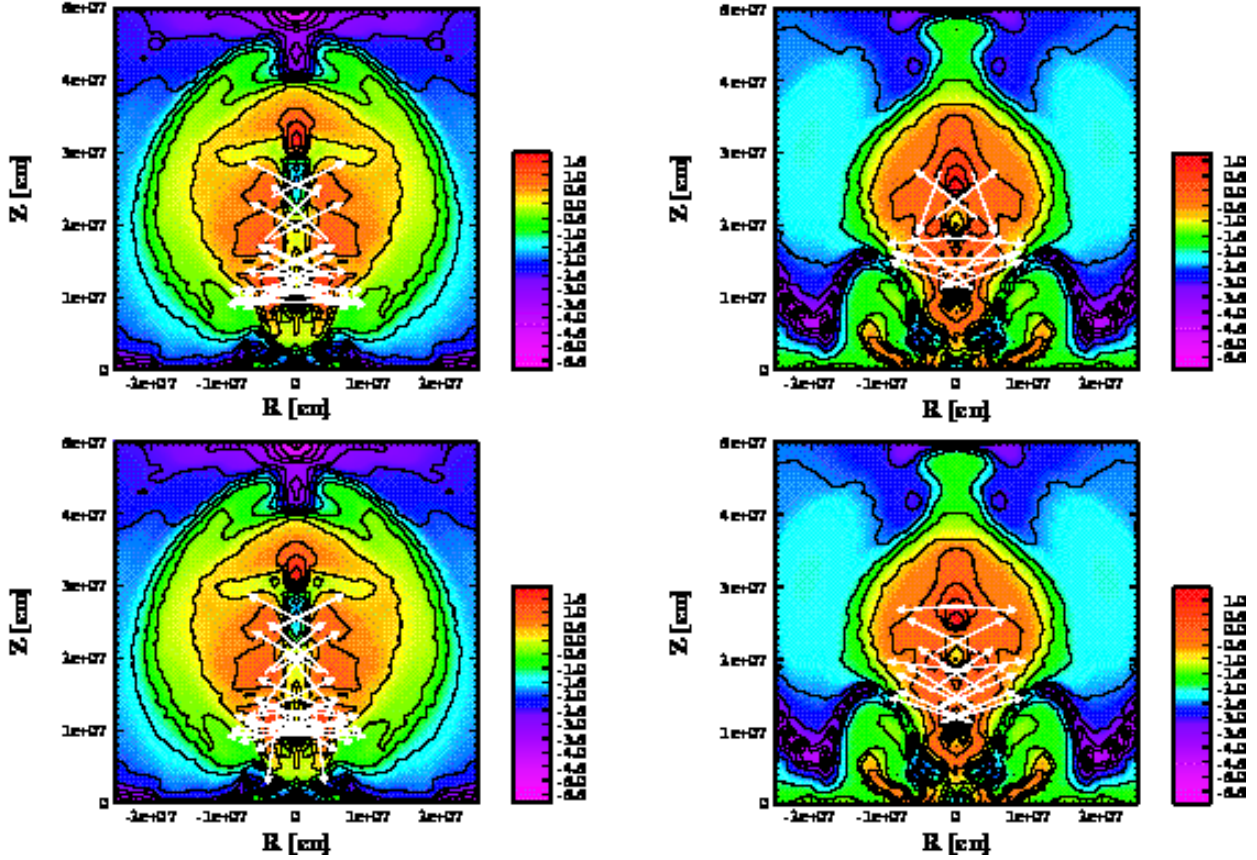} 
  \caption{The Lorentz forces (arrows) on top of the color contour of
 the ratio $p_{mag}/p_{matter}$ for model H10 at 23 ms after bounce
 (161 ms from the onset of collapse) and for model Q10 at 30 ms after
 bounce (161 ms from the onset of collapse) (top left and top
 right panels, respectively).
 The sum of the Lorentz force and matter-pressure-gradient on top of the
 same color contour for models H10 and Q10 (bottom left and bottom
 right panels,
 respectively) at the same evaluation times as for
 top figures. The shock is located at about 500 km from the center
 on the rotation axis. The arrows are drawn only where $p_{mag}$ is
 greater than $8p_{mattar}$ in the left panels and
 $\frac{1}{2}p_{mattar}$ in the right panels .}
  \label{lorentz}
\end{figure}

Next, we discuss what causes the differences in dynamics among these
models such as the asymmetry of shock
front and the jet collimation. We pay attention to the
Lorentz force which is the third term in r.h.s of Eq. (\ref{moc}),
$\frac{1}{4\pi}(\nabla\times\mbox{\boldmath $B$})\times\mbox{\boldmath
$B$}$. In the top panels of Fig. \ref{lorentz}
we show the Lorentz force for models H10 and Q10. The lower panels in
the figure show the total force field including 
the matter-pressure-gradient. It can been seen clearly in each case that
the Lorentz force thrusts matter to the narrow region around the rotation
axis. Even after the pressure gradient which tends to expand matter is
added, the total force still squeezes matter and helps to push it more
powerfully along the axis. 

One can also see the difference between the models;
the difference in the width of the magnetic-force-dominant region. This
region is wider for model H10 and more matter is forced to collimate to
form a jet. For model Q10, on the other hand, the region is narrower and
less matter is affected by the Lorentz force, which leads to the remarkable
collimation as seen in Fig. \ref{veloc}. The asymmetry of the shock
front also depends on the width of the magnetic-field-dominant
region. In the models with the
initially axially-concentrated fields, this region is narrower than the
model with the
initially uniform field, which causes the higher aspect ratio of the
shock front. The models with the initially quadrapole-like fields are
exceptions. This is because the magnetic-force-dominant region exist
also near the equatorial plane (see Fig. \ref{lorentz}). In
this region, the magnetic field rather chaotic and we have not shown
them in Fig. \ref{lorentz}. Nevertheless, the Lorentz
force on average pushes matter in the horizontal direction. 

We define the explosion energy as the sum of kinetic, internal,
gravitational and magnetic energy of the region where the sum is positive
when the shock reaches 800 km. This is admittedly a crude estimation of
the true explosion energy particularly where it is evaluated at early
times. We can still infer the
trend of the explosion strength among the models. The explosion energy
in each model depends on the initial field configuration as found in
Table \ref{parameter}.
In fact, the models with initially axially-concentrated fields result
in more energetic
explosions than the case with initially uniform field. Looking in more
detail, we find that the main differences are in
kinetic and magnetic energies. The models with ``C'' in name have
greater velocity and magnetic field and the mass of ejected matter is also
larger. In the case of quadrapole-like field, its explosion energy is
smaller than that in the case of the
uniform field by almost an order of magnitude although the velocity of
ejected matter
is quite high. This is because the mass of matter which have positive
total energy is small.

\subsection{Amplification of Magnetic Field}

\begin{figure}
\plotone{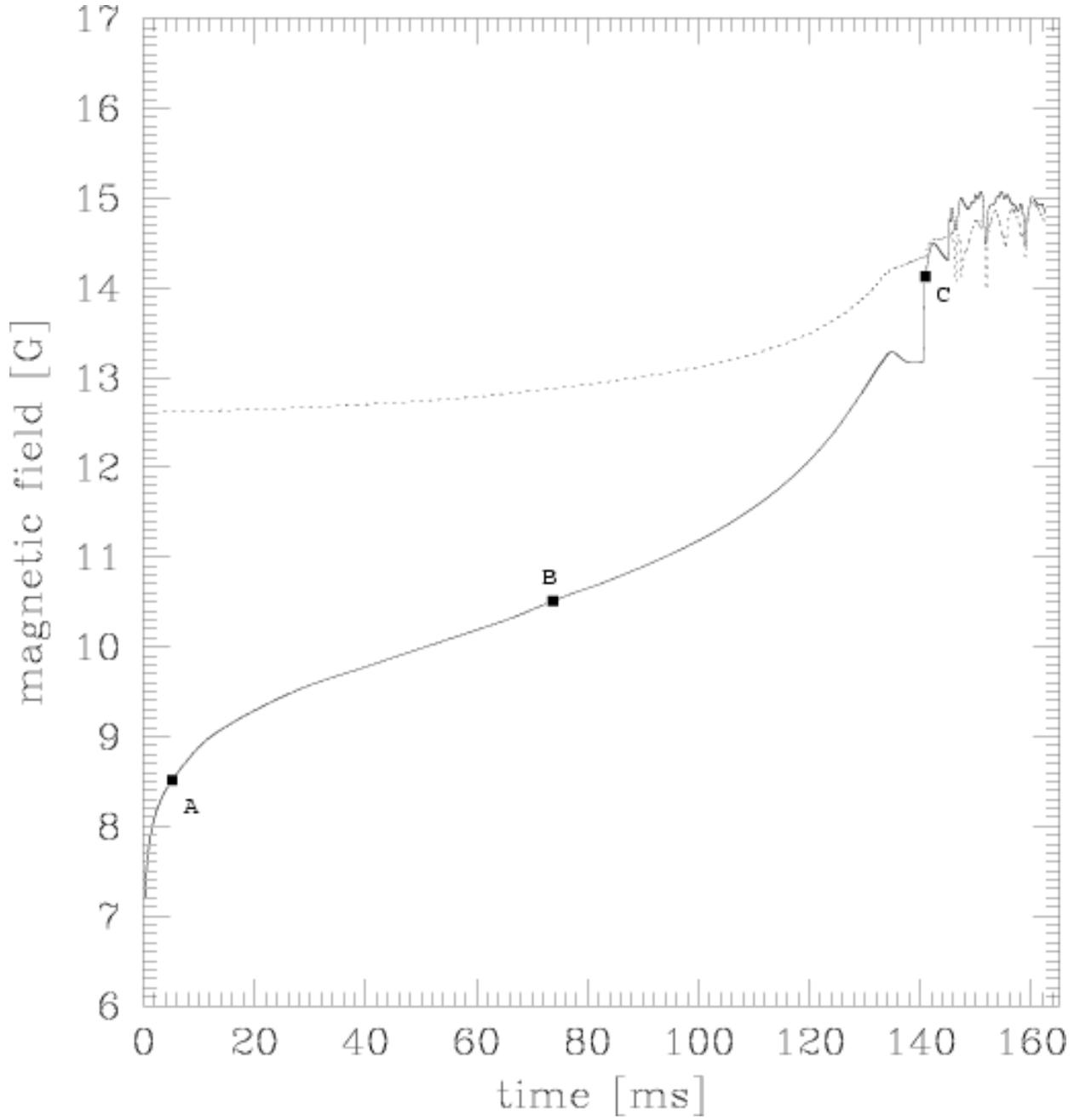}
  \caption{Time evolution of poloidal and toroidal fields at ($r$,
 $\theta$) = (100 km, $18^\circ$) for
 model H10. The dotted and solid lines correspond to the
 poloidal and toroidal components, respectively. }
  \label{btime}
\end{figure}

The amplification of magnetic field is one of the most important issues in this
study. In Fig. \ref{btime} we show the time evolutions of magnetic field for
models H10. It can be seen that the initially negligible
toroidal field is amplified up to values comparable to the poloidal field
which has also grown by about four orders of magnitude from its initial
strength. It is important to know what process amplifies
the magnetic field so
greatly.  The compression of frozen-in field can amplify the magnetic
field. The core contracts
from the initial radius, $\sim 10^{8}$ cm, to the final radius some
$10^{6}$cm. Hence, the field expected to grow by nearly four orders of
magnitude by this process alone. Other possible agencies for
field-amplification are the field-wrapping by differential rotation
\citep{mei76} and the magnetorotational instability (MRI) \citep{bal91,aki03}.

The maximum poloidal field in model H10 grows from $\sim 10^{12}$ to $\sim
10^{16}$ G. The amplitude of the poloidal field-growth, four orders of
magnitude, is common to all models, and is
just the value expected for the field-compression. This implies the
poloidal field is amplified almost entirely by the compression and it
does not seen that any instabilities play an important role in our models. 

For the toroidal field, we make an estimate for the  
field-growth rates by field-wrapping and the compression separately. The field
wrapping is a process which produces toroidal 
field from poloidal field by differential rotation. We can
estimate this rate by extracting from the $\phi$-component of r.h.s. of
Eq. (\ref{far}) the terms which include rotation velocity and
poloidal field;
\begin{equation}
\Biggl(\frac{\partial B_\phi}{\partial t}\Biggr)_{wrap}=\frac{1}{r}\Biggl[\frac{\partial (rv_\phi B_r)}{\partial r}+\frac{\partial (v_\phi B_\theta)}{\partial \theta}\Biggr]\equiv\alpha.
\label{bgw}
\end{equation}
For the compression, we extract the terms which include $v_r$,
$v_\theta$ and $B_\phi$ from the $\phi$ component of r.h.s. of
Eq. (\ref{far}) as: 
\begin{equation}
\Biggl(\frac{\partial B_\phi}{\partial t}\Biggr)_{comp}=-\frac{1}{r}\Biggl[\frac{\partial (r v_r B_\phi)}{\partial r}+\frac{\partial (v_\theta B_\phi)}{\partial \theta}\Biggr]\equiv\beta.
\label{bgf}
\end{equation}

We roughly divide the time evolutions into four phases in
terms of the toroidal field-amplification as shown in Fig. \ref{btime}. For
model H10, for example, the first phase is a period from 0 ms to about
10 ms, which shows 
a first steep gradient in the evolution of maximum fields and includes
point A. The second phase is a period from about 10 ms to 140 ms, which
corresponds to a gentle growth and contains point B. The third
phase continues from about 140 ms to 142 ms, where the second steep
gradient appears and point C is representative. This phase
corresponds to the period just after the bounce in Fig. \ref{dch10}.
The last phase is the period from about 142 ms to the end, which shows
a rather chaotic
evolution. The growth rates
$\alpha$, $\beta$ are presented in Table
\ref{bgrate}. During the first and second phases, the field-wrapping
dominates
the compression. During the third phase, on the
contrary,
the compression becomes dominant over the wrapping which, however, still plays
an important role. At the end of the third phase, the ratio of magnetic
pressure to matter pressure reaches almost unity and magnetic force
begins to play an important role. After the third phase,
the maximum toroidal field shows 
oscillations and grows very slowly. In facts, the growth rates $\alpha$
and $\beta$ also
oscillate and are responsible for the oscillating field evolutions. We stop our
computations at about 160 - 190 ms when the shock wave reaches 800 km
for exploding models.

While the above
feature of field evolution is common to all models, the final strength of
resulting fields vary from model to model. As shown in Table \ref{parameter},
the final magnetic energy as a result of amplifications is larger in models
C's and Q's than in model H10 though they have the same initial magnetic
energy. This is simply because the initial fields of models
C's and Q's are more centrally concentrated compared to model
H10. Since the field
amplification occurs in the vicinity of the boundary between the inner and
outer cores, concentrated fields are advantageous. This then causes more
energetic explosions in the initially axially-concentrated
field models. In the quadrapole-like field models, the explosion energy
is rather small since the amplification region is smaller and contains smaller
mass.

\subsection{Models with Initially Weak Fields} 
Models with initially weak fields do not lead to substantial matter
ejection no matter what the initial configuration is adopted. In fact, there
appears no region where magnetic pressure is comparable to matter
pressure through the whole simulations. It seems that MRI-like
instability do not occur in our models,
which would lead to the exponential growth of fields. As a result,
magnetic fields do not evolve large enough to affect dynamics. 

\section{Discussion and Conclusion}
We have done two-dimensional MHD simulations on core-collapse of
massive stars for several initial configurations of magnetic field, which
include the uniform and axially-concentrated fields, quadrapole-like
field and toroidal fields. Different
differential rotations have been considered. We mainly focus on the
systematic trends among the models. 

Since it is impossible at the moment to observe magnetic fields of
presupernova stars, we have treated the field strength and
configuration as free parameters in this study. Nevertheless, there are
some suggestions from the theoretical studies of stellar evolutions
\citep{heg03} although they are admittedly uncertain. For example, the toroidal
field may be dominant prior to core-collapse, since the differential rotation
is inevitably produced as the core contracts in the quasi-static
evolutionary stages. It may be also possible that magnetic field may be
centrally concentrated if it traces the density profile. Bearing these
suggestions in mind, we have studied the effect of
poloidal fields and toroidal fields separately and have also adopted
centrally-concentrated field configurations.

In the models with axially-concentrated fields the explosion energy is
about 1.5 times
as large as that of the model with uniform field if the initial ratio of
magnetic
energy to gravitational energy is identical. Although we have not
calculated a model 
with more small values of $r_0$ and $\varpi_0$, it is expected that
larger degrees of field concentration or differential rotation will give
more energetic explosions. According to \citet{mei76},
there may exist a threshold in the initial strength of magnetic field
for the MHD
explosion. We expect that the parallel field concentration will lower
the threshold if any at all. 

Our models with quadrapole-like fields give tightly
collimated and very high
velocity but less energetic explosions. The velocity becomes about half
the light speed. This result might have some implications for gamma-ray
bursts (GRB). In the paper of \citet{whe02}, they argued a scenario, in which
a ``fast
toroidal jet'' is produced first but does not lead to the MHD explosion,
and a black hole is formed. Then the second even faster, highly
relativistic jet is supposed to be produced from
the black hole and interacts with the first slower jet
to give GRB. The jets in model Q's might be something like ``fast
toroidal jet'' in
their scenario, since our jet is fast and toroidal-field dominant, and
carries only a small part of core mass so that there would remain
plenty of energy to produce the second highly
relativistic jet. Our jet will be able to sweep out baryons on its way,
which will then help the second jet be accelerated to an extremely
relativistic velocity. Note that, however, their ``fast
toroidal jet'' is generated in proto-pulsar phase and the origin is
different from our jet. The present mesh resolution is not sufficient to
properly treat a tight collimation, and numerical simulations with finer
angular mesh are needed. Special relativity should be also included.    

Our results for the quadrapole-like fields look quite different from that of 
\citet{ard04} in which the almost same field configuration is employed. Their
simulation resulted in more energetic explosion,
$0.6\times10^{51}$erg, than ours and matter is ejected more strongly in the
direction parallel to the equatorial plane. They employed a 2D implicit
Lagrangian code with a parametric EOS,
neutrino losses and iron dissociations by introducing
approximate formulae. The main difference is the way to set the initial
magnetic field. They
'turned on' the quadrapole-like field well after the collapse of the core,
when the toroidal fields have already grown in our models. What is more,
as they set the initial ratio of magnetic energy to gravitational energy
to be $10^{-6}$, the growth of magnetic field is further delayed. We
think this is the reason for apparent difference.

\citet{aki03} claimed that MRI is likely to grow in the postbounce core
\citep[see, however, ][]{fry04}. In the present simulations, as
mentioned already MRI-like field-amplification process is not found. In
particular, the initial weak fields in models H10w, C5w, Q5w, T5w
do not develop so mach as to influence dynamics. This is again at odds
with the results of \citet{ard04}. We suppose that this is mainly due to
the spacial resolution of our simulations. It should be note that since
the toroidal fields are highly likely to be dominant in the post bounce
core and MRI operates also non-axisymmerically, 3D simulations will be
important in considering MRI. We are already undertaking three dimensional MHD
simulations with high resolution scheme \citep{saw05}, and results will
be presented elsewhere. 

If MRI grows efficiently form weak seed fields as some authors claimed,
the normal supernova may be also explained
by the MHD process \citep{whe02}. Then we may have to worry about how to
reduce the magnetic fields by the time the pulsar is observed, since
otherwise the magnetar would be a common product. These are important
issues for the future work.

\acknowledgments
Some of the numerical simulations were done on the
supercomputer VPP700E/128 at RIKEN and VPP500/80 at KEK (KEK
Supercomputer Projects No. 108). This work was partially supported by
the Grants-in-Aid for the Scientific 
Research (14740166, 14079202) from Ministry of Education, Science and
Culture of Japan and by the Grants-in-Aid for the 21th century COE
program ``Holistic Research and Education Center for Physics of
Self-organizing Systems''.

\begin{table}
\begin{center}
\caption{Magnetic Field and Differential Rotation for Initial Models}\label{models}
\begin{tabular}{lcccccc}
\tableline\tableline
Model & $|E_m/W|$ [\%] & $|T/W|$ [\%]& $B_i$ [G] & $\Omega_i$ [rad/s] &
 $r_0$ \textit{or} $\varpi_0$ [km] & $R_0$ [km] \\
\tableline
R10 & 0.0 & 0.5 & 0.0 & 0.0 & - & - \\
H10 & 0.5 & 0.5 & $4.1\times10^{12}$ & 3.9 & $\infty$ & 1000 \\
C5  & 0.5 & 0.5 & $1.4\times10^{13}$ & 7.0 & 500  & 500\\
C10 & 0.5 & 0.5 & $8.2\times10^{12}$ & 3.9 & 1000 & 1000\\
Q3  & 0.5 & 0.5 & $8.4\times10^{12}$ & $1.3\times10$ & - & 300\\
Q5  & 0.5 & 0.5 & $8.4\times10^{12}$ & 7.0 & - & 500\\
Q10 & 0.5 & 0.5 & $8.4\times10^{12}$ & 3.9 & - & 1000\\
T3  & 0.5 & 0.5 & $5.1\times10^{13}$ & $1.3\times10$ & 300  & 300\\
T5  & 0.5 & 0.5 & $2.6\times10^{13}$ & 7.0 & 500  & 500\\
T10 & 0.5 & 0.5 & $1.1\times10^{13}$ & 3.9 & 1000 & 1000\\
H10w & 10$^{-8}$ & 0.5 & $5.8\times10^{8}$ & 3.9 & $\infty$ & 1000\\
C5w & 10$^{-8}$ & 0.5 & $2.0\times10^{9}$ & 7.0 & 500 & 500\\
Q5w & 10$^{-8}$ & 0.5 & $1.2\times10^{9}$ & 7.0 & - & 500\\
T5w & 10$^{-8}$ & 0.5 & $3.6\times10^{9}$ & 7.0 & 500 & 500\\
\tableline
\end{tabular}
\tablecomments{$|E_m/W|$ : the magnetic energy normalized by the
 gravitational energy. $|T/W|$ : the rotation energy normalized by the
 gravitational energy. $B_i$ : the initial maximum magnetic field. $\Omega_i$ : the initial angular velocity at the center of
 the core. $r_0$ and $\varpi_0$ : the parameters which specify the
 degree of field
 concentration (see Eqs. (\ref{con}) and (\ref{tro})). $R_0$ :
 the parameter which specifies the degree of the differential rotation (see
 Eq. (\ref{dif})).}
\end{center}
\end{table}

\begin{deluxetable}{lcccccccccccccc}
\tabletypesize{\tiny}
\rotate
\tablecaption{Key Parameters for All Models}\label{parameter}
\tablewidth{0pt}
\tablehead{
\colhead{Model} & \colhead{$M_{IC}$} &
 \colhead{$|E_m/W|_b$} &
\colhead{$|T/W|_b$} & \colhead{$B_b,max$} & 
\colhead{$\Omega_b,max$} & \colhead{$|E_m/W|_{fin}$} & \colhead{$|T/W|_{fin}$} &
\colhead{$B_{fin,max}$} & \colhead{$\Omega_{fin,max}$} &
\colhead{$v_{fin,max}$} & \colhead{$E_{exp}$} & \colhead{$M_{ej}$} &
 $\Delta\theta$ & $f_{shock}$}

\startdata
R10 & 0.82 & 0.0 & $1.0\times10$ & 0.0 &
 $7.9\times10^3$ & 0.0 & 6.8 & 0.0 & $7.5\times10^3$ & $4.9\times10^8$ &
 0.12 & 0.013 & 180 & 0.98 \\ 
H10 & 0.82 & 0.15 & 8.9 & $2.0\times10^{16}$ &
 $1.0\times10^4$ & 0.62 & 4.0 & $2.2\times10^{16}$&
 $5.3\times10^3$ & $4.6\times10^9$ & 1.6 & 0.18 & 42 & 1.5 \\
C5  & 0.82 & 0.70 & 9.1 & $1.6\times10^{16}$ &
 $8.0\times10^3$ & 1.3 & 4.4 & $6.5\times10^{16}$ & $8.7\times10^{3}$ &
 $6.3\times10^{9}$ & 2.1 & 0.16 & 42 & 2.0 \\
C10 & 0.81 & 0.25 & 8.0 &
 $1.8\times10^{16}$ & $7.3\times10^3$ & 1.1 & 3.7 & $6.6\times10^{16}$ &
 $3.3\times10^3$ & $6.4\times10^9$ & 2.3 & 0.17 & 54 & 1.9 \\
Q3  & 0.89 & 0.075 & $1.1\times10$ &
 $1.0\times10^{16}$ & $9.8\times10^{3}$ & 0.90 & 6.8 &
 $2.6\times10^{16}$ & $8.3\times10^3$ & $9.8\times10^{9}$ & 0.24 & 0.021
 & 6 & 1.6\\
Q5  & 0.83 & 0.060 & $1.0\times10$ &
 $7.0\times10^{15}$ & $8.0\times10^3$ & 0.96 & 5.9 & $3.0\times10^{16}$
 & $8.5\times10^3$ & $5.2\times10^{9}$ & 0.26 & 0.029
 & 12 & 1.4 \\
Q10 & 0.81 & 0.041 & 9.0 & $1.0\times10^{16}$ &
 $1.1\times10^4$ & 1.1 & 5.0 & $6.5\times10^{16}$ & $6.8\times10^3$&
 $1.3\times10^{10}$ & 0.59  & 0.066 & 24 & 1.3 \\ 
T3  & 0.91 & 0.58 & $1.0\times10$ &
 $9.0\times10^{16}$ & $4.2\times10^3$ & 0.52 & 7.2 & $3.3\times10^{16}$
 & $2.6\times10^4$ & $4.8\times10^8$ & 0.11 & 0.012
 & 180 & 0.93 \\
T5  & 0.82 & 0.33 & $1.0\times10$ &
 $7.1\times10^{16}$ & $2.7\times10^4$ & 0.30 & 6.8 & $1.8\times10^{16}$
 & $9.3\times10^{3}$ & $5.4\times10^8$ & 0.091 & 0.010 & 180 & 0.89 \\
T10 & 0.81 & 0.12 & 8.4 & $5.7\times10^{16}$ &
 $1.8\times10^4$ & 0.11 & 5.9 & $1.4\times10^{16}$ & $1.1\times10^4$ &
 $6.5\times10^8$ & 0.14 & 0.016 & 180 & 0.93 \\
H5w & 0.80 & $2.3\times10^{-9}$ & 8.2 &
 $3.5\times10^{12}$ & $8.6\times10^3$ & $5.2\times10^{-7}$ & 6.1 &
 $5.5\times10^{13}$ & $1.2\times10^4$ & $8.8\times10^8$ & 0.14 & 0.027 &
 180 & 1.2 \\
C5w & 0.82 & $3.2\times10^{-8}$ & $1.0\times10$ &
 $1.0\times10^{13}$ & $7.9\times10^3$ & $5.6\times10^{-6}$ & 6.9 &
 $5.4\times10^{13}$ & $9.0\times10^3$ & $8.2\times10^8$ & 0.11 & 0.018 &
 180 & 0.95 \\
Q5w & 0.82 & $1.4\times10^{-9}$ & $1.0\times10$ &
 $1.5\times10^{12}$ & $7.9\times10^{3}$ & $3.0\times10^{-7}$ & 6.9 &
 $6.2\times10^{12}$ & $9.2\times10^{3}$ & $8.0\times10^{8}$ & 0.10 &
 0.017 & 180 & 0.97\\
T5w & 0.82 & $6.8\times10^{-9}$ & $1.0\times10$ &
 $1.9\times10^{13}$ & $7.9\times10^{3}$ &
 $6.4\times10^{-9}$ & 6.9 & $2.7\times10^{12}$ & $9.2\times10^{3}$ &
 $7.9\times10^{8}$ & 0.10 & 0.017 & 180 & 0.97 \\
\enddata

\tablecomments{$M_{IC}$ : the inner core mass at bounce in unit of
 $M_\odot$. $\rho_c$ : the central density at bounce in
 $10^{14}$ g/cm$^3$. The ratios $|E_m/W|$ and $|T/W|$ are given in
 percentage. $B_{max}$ : the maximum magnetic field in
 G. $\Omega_{max}$ : the maximum angular velocity in rad/s. $v_{max}$
 : the maximum positive radial velocity in cm/s. $E_{exp}$ : the
 explosion energy in $10^{51}$erg. $M_{ej}$ : the ejected mass in
 $M_\odot$. $\Delta \theta$ : the opening angle of the induced jet in
 degree (see the footnote in \S \ref{result} for the definition). $f_{shock}$
 : the aspect ratio. For all parameters, the subscripts
 ``$b$'' and ``$fin$''
 denote the values at bounce and those at the end of calculations,
 respectively.}
\end{deluxetable}

\begin{table}
\caption{The Field Growth Rate in Model H10}\label{bgrate}
\begin{tabular}{ccccc}
\tableline\tableline
Point & $t_{eva}$ [ms] & $|B_{\phi}|$ [G] & $|\alpha|$ [G/s] & $|\beta|$ [G/s]\\
\tableline
A & 5.2 & $3.2\times10^{8}$ & $5\times10^{11}$ & $8\times10^{8}$\\
B & 73.7 & $3.2\times10^{10}$ & $2\times10^{12}$ & $3\times10^{11}$\\
C & 140.9 & $1.3\times10^{14}$ & $4\times10^{16}$ & $1\times10^{17}$ \\

\tableline
\end{tabular}
\tablecomments{Each point is shown in Fig. \ref{btime}. $t_{eva}$ : time
 at each
 point (evaluation time). $|B_{\phi}|$ : the toroidal field at
 $t_{eva}$. $\alpha$ : the field-growth rate by field-wrapping
 (see Eq. (\ref{bgw})). $\beta$ : the field-growth rate by
 compression (see Eq. (\ref{bgf})).} 
\end{table}

\end{document}